\documentclass[a4paper]{jpconf}

\newcommand{\be}{\begin{equation}}
\newcommand{\ee}{\end{equation}}
\newcommand{\bea}{\begin{eqnarray}}
\newcommand{\eea}{\end{eqnarray}}
\newcommand{\ben}{\begin{eqnarray*}}
\newcommand{\een}{\end{eqnarray*}}
\newcommand{\doa}{\downarrow}
\newcommand{\upa}{\uparrow}
\newcommand{\size}{50}

\usepackage{graphicx}
\usepackage{amssymb}
\usepackage{amsmath}
\usepackage{amssymb}
\usepackage{epsfig}
\usepackage[dvips]{color}
\usepackage{url}
\begin{document}

\title{Correlation-induced band suppression in the two-orbital Hubbard model}

\author{E Plekhanov$^{1,2}$, A Avella$^{1,2,3}$, F Mancini$^{1,2}$ and F P Mancini$^{1,3,4}$}

\address{$^1$ Dipartimento di Fisica ``E.R. Caianiello'', Universit\`a
degli Studi di Salerno, 84084 Fisciano (SA), Italy}

\address{$^2$ Unit\`a CNISM di Salerno, Universit\`a degli Studi di
Salerno, 84084 Fisciano (SA), Italy}

\address{$^3$ CNR-SPIN, UoS di Salerno, 84084 Fisciano (SA), Italy}

\address{$^4$ Sezione INFN di Perugia, 06123 Perugia, Italy}

\ead{plekhanoff@physics.unisa.it}

\begin{abstract}

The orbital degrees of freedom are of vital importance in explanation of
various phenomena. Among them is the orbital-selective Mott transition
(OSMT), which is thought to occur in several materials as
Ca$_{2-x}$Sr$_x$RuO$_4$ and La$_{n+1}$Ni${_n}$O$_{3n+1}$. OSMT is
usually studied in the infinite-dimension limit, and for the time being,
it is not clear if it could survive in one-dimensional (1D) case.
There exist two scenarios for the OSMT: upon increasing the interaction
in a two-band system i) one of the bands becomes insulating, while the
other remains metallic and ii) one of the bands becomes empty, while the
other may eventually undergo a single-band Mott insulator transition.
In this work, we present a preliminary study of the two-orbital Hubbard
model by means of Density Matrix Renormalization Group in 1D at
quarter-filling, where the second scenario seems to be realized.
In particular, we study the orbital densities, double occupancies and
form-factors also in the case of finite inter-orbital inter-site hopping.

\end{abstract}

%
%
\section{Introduction}

The crucial role of the orbital degrees of freedom in explaining several
intriguing phenomena such as metal-insulator transition and magnetism in
doped phthalocyanines as well as the superconductivity in heavy-fermion
compounds has been recently recognized. On the theoretical side, the
study of multi-orbital systems is extremely complicated and has been
mainly accomplished in the infinite-dimension limit by means of the
Dynamical Mean-Field Theory, although, in these systems, the spatial
correlations are essential.

Another fascinating phenomenon inherent to the multi-orbital systems is
the orbital selective Mott transition (OSMT). In such a transition, by
tuning the strength of the Coulomb repulsion, it is possible to open a
gap in some band(s), while leaving the other(s) gapless. It is thought
that OSMT occurs in materials as
Ca$_{2-x}$Sr$_x$RuO$_4$~\cite{nakatsuji_01,nakatsuji_02} and
La$_{n+1}$Ni${_n}$O$_{3n+1}$~\cite{sreedhar,zhang,kobayashi}.

The simplest multi-orbital model is the two-orbital Hubbard model, whose
most general Hamiltonian reads as follows:

\bea
   H &=& -\sum_{\langle ij \rangle\alpha\beta\sigma} t^{\alpha\beta}
   \left(
   c^{\dagger}_{i\alpha\sigma}c^{\phantom{\dagger}}_{j\beta\sigma} + h.c. \right)
    + U \sum_{i\alpha} n_{i\alpha\upa}n_{i\alpha\doa}\\
   \nonumber
   &+& \left( U^{\prime} - \frac{J}{2} \right)
   \sum_{i} \left( n_{i1\upa} + n_{i1\doa} \right) \left( n_{i2\upa} 
   + n_{i2\doa} \right)
   -2J\sum_{i}\sum_{k=1}^{3} S^{k}_{i1} S^{k}_{i2}
   + J \sum_{i} \left[ p^{\phantom{\dagger}}_{i1} p^{\dagger}_{i2}
   + p^{\phantom{\dagger}}_{i2} p^{\dagger}_{i1} \right].
   \label{ham}
\eea

Here $t^{\alpha\beta}$ parametrizes intra- ($t^{11}$ and $t^{22}$) and
inter- orbital ($t^{12}$) hoppings between nearest-neighbor (NN) sites,
$U$ and $U^{\prime}$ are the on-site inter- and intra-orbital Coulomb
repulsions, respectively, while $J$ is the direct exchange interaction.
Rotational symmetry imposes that $U=U^{\prime}+2J$. Finally,
$c^{\dagger}_{i\alpha\sigma}$ creates an electron with spin $\sigma$
in the $\alpha$-orbital of site $i$,
$S^{k}_{i\alpha}\equiv\frac{1}{2}\sum
\sigma^{k}_{ab}c^{\dagger}_{i\alpha a}c^{\phantom{\dagger}}_{i\alpha b}$
is the electron spin operator in the $\alpha$-orbital of site $i$, being
$\sigma^{k}_{ab}$ the Pauli matrices, and the pair operator
$p_{i\alpha}$ and the electron density operator $n_{i\alpha\sigma}$ are defined as follows: 
$n_{i\alpha\sigma}=c^{\dagger}_{i\alpha\sigma}c^{\phantom{\dagger}}_{i\alpha\sigma}$,
$p_{i\alpha}=c_{i\alpha \upa}c_{i\alpha \doa}$.

In the present manuscript, we use finite-size Density Matrix
Renormalization Group (DMRG) method~\cite{white} to investigate the
zero-temperature properties of the Hamiltonian~(\ref{ham}). We consider
chains of $\size$ sites with open boundary conditions. In the DMRG
decimation procedure, we retain up to $M=300$ lowest eigenstates of the
density matrix, which amounts to have a truncation error on the sum of
the density matrix eigenstates not larger that $10^{-5}$.
\begin{figure*}
  \begin{tabular}{cc}
      \epsfig{file=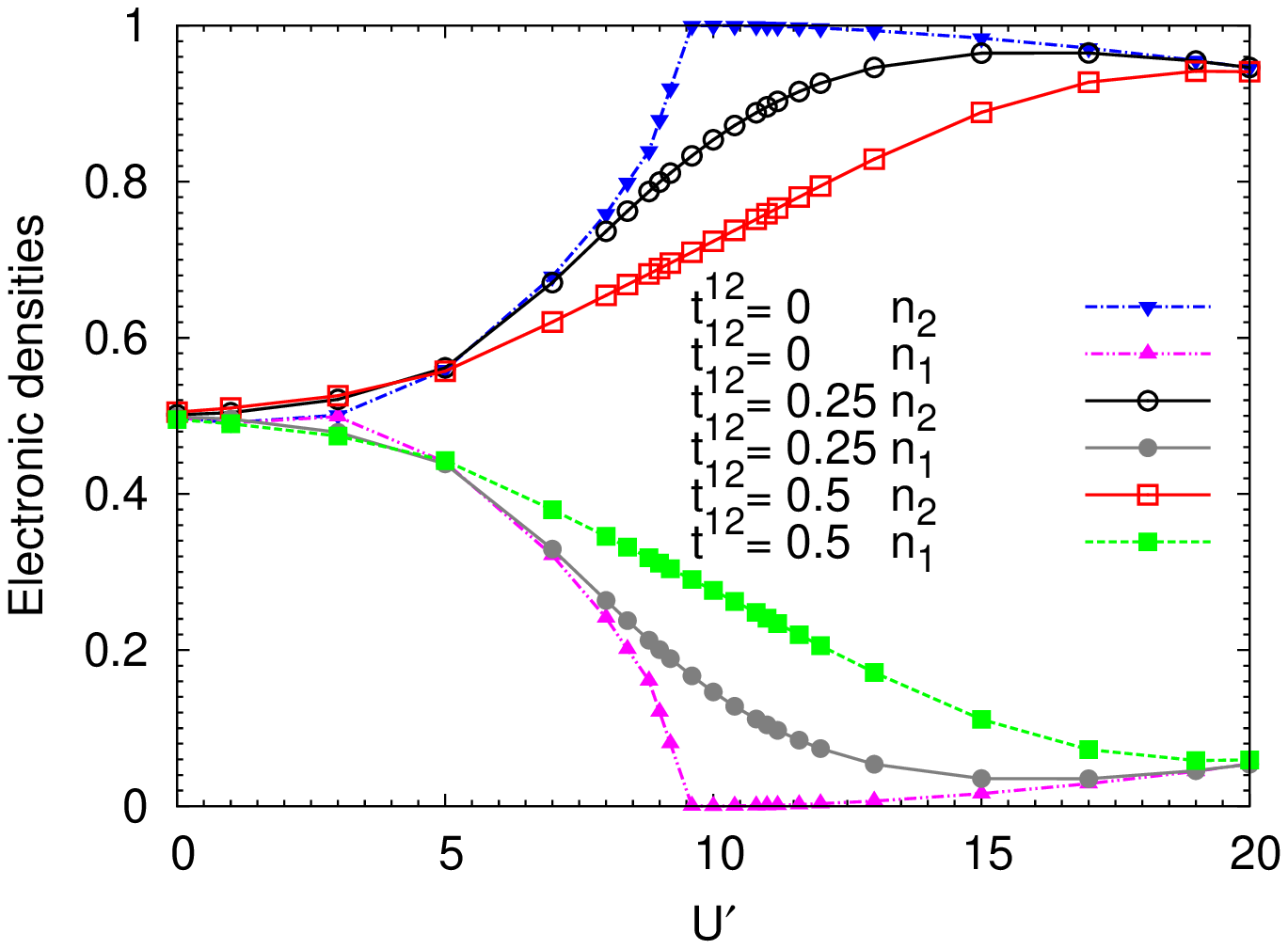, angle=270,scale=0.31}
      \epsfig{file=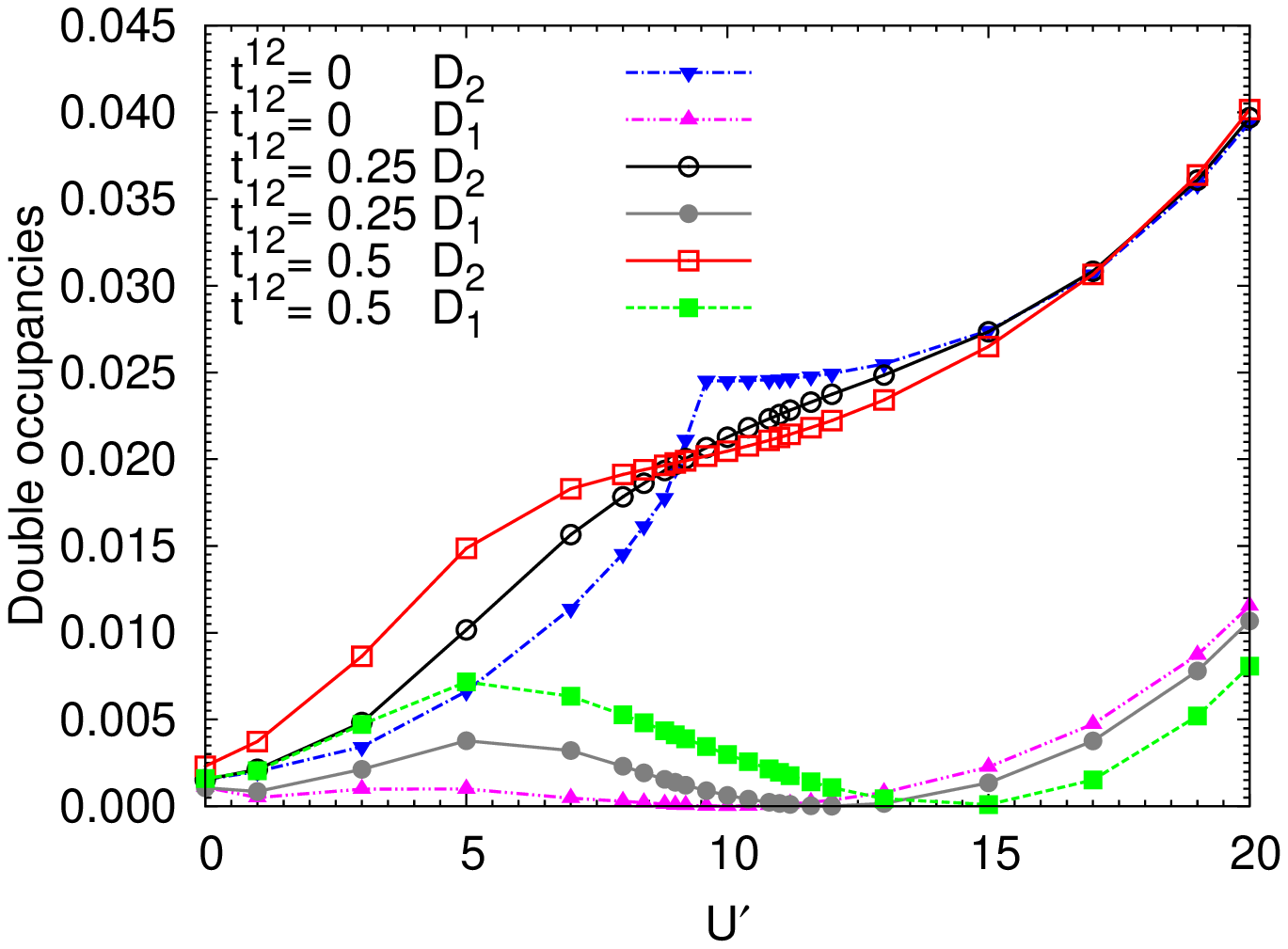 , angle=270,scale=0.31}
  \end{tabular}
   \caption{Left panel: $U^{\prime}$-dependence of the average occupation number
   per orbital per site $n_1$ and $n_2$ at different values of
   inter-orbital hopping $t^{12}$. $U=10$, $t^{11}/t^{22}=0.5$, DMRG
   $\size$ sites. Right panel: The same for the average double occupancy
   $D_1$ and $D_2$.
   }
   \label{fig1}
\end{figure*}
\begin{figure*}
	\begin{center}
      \epsfig{file=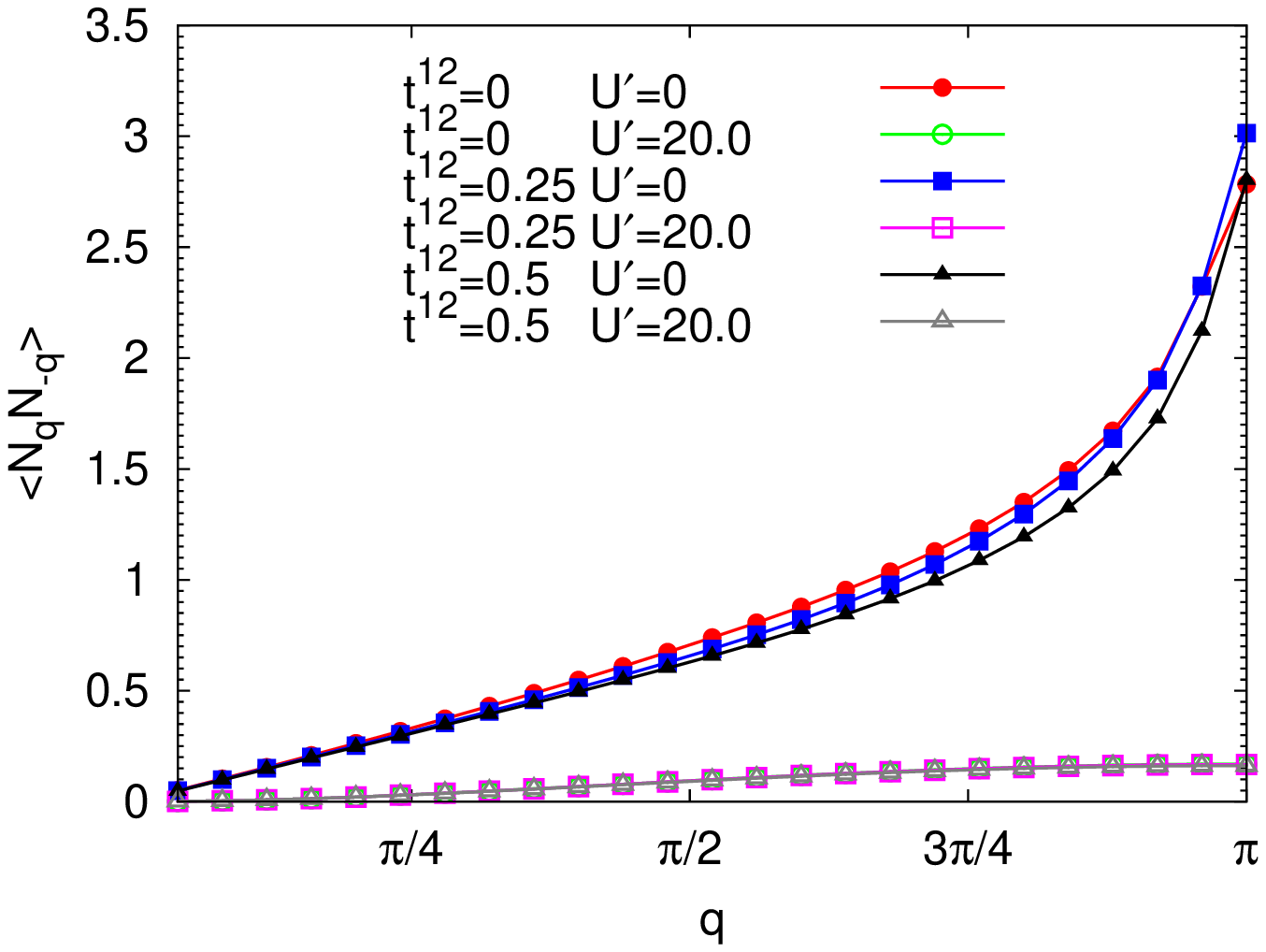, angle=270,scale=0.31}
      \epsfig{file=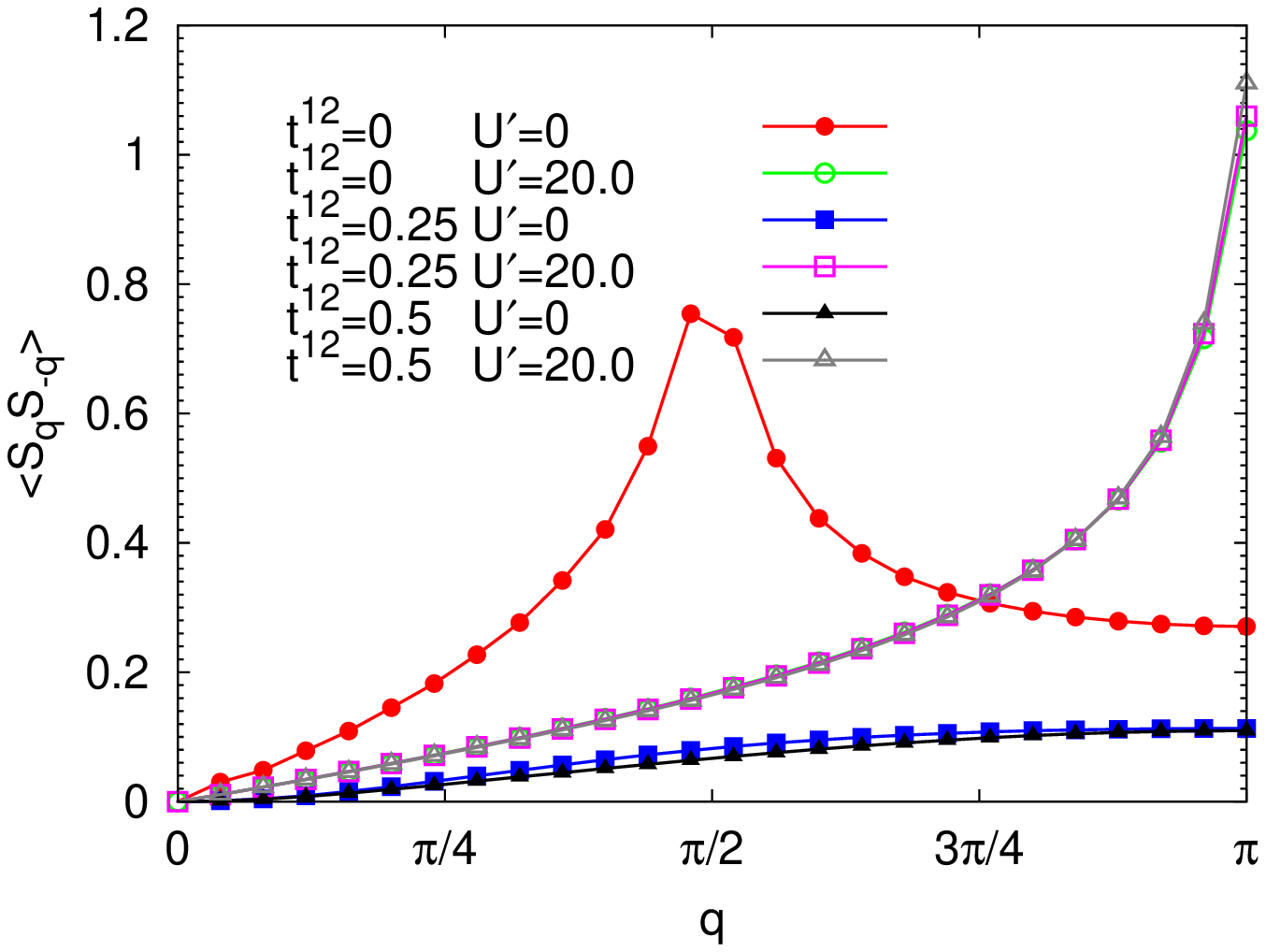, angle=270,scale=0.31}
	\end{center}
   \caption{Fourier transforms of the density (Left panel) and spin
   (Right panel) static form-factors at initial ($U^{\prime}=0$) and
   final ($U^{\prime}=20$) points from Fig.~\ref{fig1} and different
   values of inter-orbital hopping $t^{12}$: $t^{12}=0,\; 0.25,\; 0.5$.
   $U=10$, $t^{11}/t^{22}=0.5$, DMRG $\size$ sites.
   }
   \label{fig2}
\end{figure*}
\section{Results}
It is well known that the (OSMT)
occurs at commensurate fillings ($n=1$, $n=2$ electrons per site). In
particular, in one-dimensional (1D) case, by means of the
product-wave-function renormalization-group technique, the existence of
two Mott states in the Hamiltonian~(\ref{ham}) at quarter- and half-
filling for $t^{12}=0$, $U=10$ and $t^{11}/t^{22}=0.5$ has been shown in
Ref.~\cite{miyashita}. It was found therein that the quarter-filling
case corresponds to a complete narrow orbital suppression (NOS), while
in the half-filling case there appears a true OSMT state. In the present
manuscript we concentrate on the case of a quarter-filled system ($n=1$)
and explore the evolution of NOS upon introducing a finite
inter-orbital hopping $t^{12}$. We have also studied the $t^{12}=0$
case, already analyzed in Ref.~\cite{miyashita}, in
order to compare our finite-size algorithm with their bulk one, as well
as to have our own $t^{12}=0$ reference line. In Fig.~\ref{fig1}, we
present our results for average electron densities and double
occupancies per orbital as functions of $U^{\prime}$ at $U=10$ and
$J=(U-U^{\prime})/2$. For $t^{12}=0$, NOS occurs approximately at
$U^{\prime}_{c}=9.4$ in qualitative agreement with
Ref.~\cite{miyashita}, although in the region of
$U^{\prime}>U^{\prime}_c$ ($J<0$) there is a slight re-population of the
otherwise empty narrow orbital. Such a re-population might be explained
as the effect of the negative $J$ Hund's exchange: the system gains more
energy by populating both orbitals rather than emptying one of them. We
have checked, by increasing $M$ (not shown), that this re-population is
not a DMRG artefact, and the small discrepancy between our results and
those of Ref.~\cite{miyashita} remains to be investigated. Our analysis
confirm the conclusion of Ref.~\cite{miyashita} that the
quarter-filled transition is due to NOS.
NOS influences inevitably the behavior of the average double occupancy per
site $D_i$ ($i=1,2$), as shown in the right panel of Fig.\ref{fig1}. As
$U^{\prime}$ grows for $U^{\prime}<U^{\prime}_c$, $D_1$ decreases
monotonically due to the depletion of the narrow orbital. At the same time,
$D_2$ grows due to the opposite reason and reaches the value $D_2\approx
0.02455$ at $U^{\prime}_c$ which is very close to the
Bethe-ansatz value ($D_{BA}=0.025$ at half-filling and $U=10$,
Ref.\cite{1DHM}).

In order to have a deeper insight into the nature of the initial and
final states involved in this transition, we have measured the static
density and spin form-factors. As shown in Fig.\ref{fig2}, in the
initial state ($U^{\prime}=0$), there is a strong density correlation
peaked at the wave-vector $\pi$, meaning that the density has staggered
correlations. These staggered correlations are gradually suppressed at
$U^{\prime}>U^{\prime}_c$. In such a strong-coupling regime, the wide
band becomes uniformly occupied leading to an almost featureless density
form-factor. The opposite picture is observed in the spin channel. At
$U^{\prime}=0$ and $t^{12}=0$, the spin form-factor exhibits a peak at
$\pi/2$, implying the enhanced staggered correlations with the period of
four lattice spacings, while at $t^{12}\ne 0$ the spin correlations are
heavily suppressed. In the strong-coupling regime at
$U^{\prime}>U^{\prime}_c$, only the wide orbital is active and the
system resembles very much a single-band Hubbard model at half-filling.
In such a case, large on-site repulsion induces strong antiferromagnetic
NN correlations, as can be seen from Fig.\ref{fig2}.
\begin{figure*}
	\begin{center}
      \epsfig{file=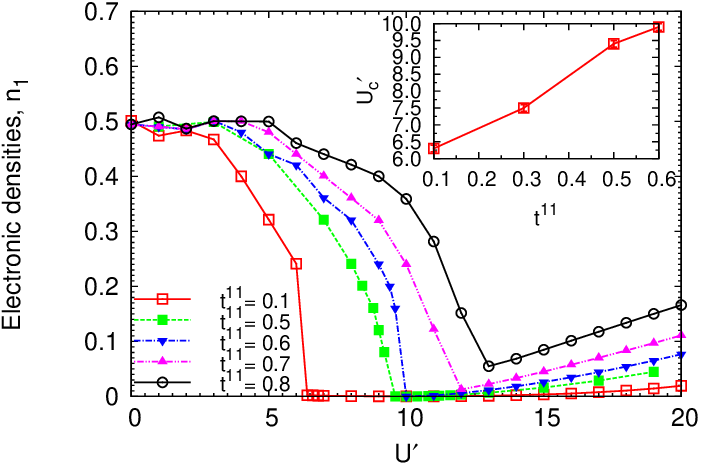, angle=0,scale=0.65}
      \epsfig{file=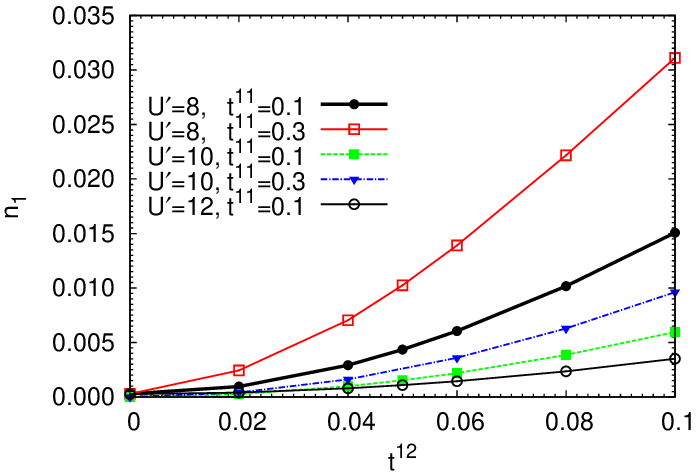, angle=0,scale=0.65}
	\end{center}
   \caption{Left panel: $U^{\prime}$-dependence of the narrow band
   occupation at $t^{12}=0$ and different values of $t^{11}$. In the
   inset, $U^{\prime}_c$ is shown as function of $t^{11}$.
   Right panel: The narrow band occupation as function of $t^{12}$ as the latter goes
   to zero at several values of $U^{\prime}$ and $t^{11}$.
   }
   \label{fig3}
\end{figure*}

Upon switching on a finite inter-orbital hopping $t^{12}$, the NOS
becomes incomplete, as shown in the left panel of Fig.~\ref{fig1}. This
can be understood by noting that the electrons in the narrow orbital can
now reach the NN site also by making the virtual hopping through the
wide orbital according to the following path:
$(i,1)\to(i+1,2)\to(i,2)\to(i+1,1)$, where $(i,\alpha)$ stands for
$\alpha$-orbital on site $i$. Such an extra possibility effectively
widens the narrow orbital. To better understand this widening, we have
investigated the NOS when $t^{12}=0$ and at different
values of $t^{11}$, as shown in the left panel of Fig.~\ref{fig3}. We
have observed the full NOS for $t^{11}\lesssim 0.6$. This full
NOS occurs at some $U^{\prime}_c(t^{11})$ (shown in the inset) above
which $n_1$ starts growing nearly linear. This situation changes
drastically when $t^{12}\ne 0$. The NOS in this case is never complete
even for small $t^{12}$, as shown in Fig.~\ref{fig1}. In the right panel
of Fig.~\ref{fig3} we present a study of the narrow band occupation as
$t^{12}\to 0$ for different values of $U^{\prime}$ and $t^{11}$. We have
chosen these values in such a way that there is a complete NOS at
$t^{12}=0$. It can be seen from Fig.~\ref{fig3} that as soon as $t^{12}$
is finite, $n_1$ is never suppressed completely.

Summarizing, by using finite-system DMRG, we have studied the influence
of the inter-orbital hopping $t^{12}$ on NOS in quarter-filled
two-orbital Hubbard model~(\ref{ham}). We have found that $t^{12}$
prevents the narrow orbital from complete suppression, by providing an
additional contribution to the intra-orbital hopping. We have found
that at $t^{12}=0$, a complete NOS occurs at $U^{\prime}_c(t^{11})$ for
$t^{11}\lesssim 0.6$ while for greater values of $t^{11}$ NOS is
incomplete for all values of $U^{\prime}\in[0,20]$. On the contrary, our
data suggest that a finite $t^{12}$ prevents complete NOS even in case of
very narrow band ($t^{11}=0.1$). In the regime with
$U^{\prime}<U^{\prime}_c$ we find a strong enhancement of staggered
charge correlations, while in the case when $U^{\prime}>U^{\prime}_c$
the staggered spin correlations are favoured.
\section{Acknowledgements}
The numerical calculations reported in the present article were done on
the CASPUR cluster (project No. $208/10$).

\end{document}